\documentclass[twocolumn, showpacs,preprintnumbers,amsmath,amssymb]{revtex4-1}
\usepackage{graphicx}
\usepackage{dcolumn}
\usepackage{bm}
\usepackage{xcolor,soul}
\usepackage{multirow}
\newcommand{\beq}{\begin{equation}}
\newcommand{\eeq}{\end{equation}}
\newcommand{\beqa}{\begin{eqnarray}}
\newcommand{\eeqa}{\end{eqnarray}}

\sethlcolor{yellow}
\setulcolor{red}
\usepackage{amsfonts}
\usepackage{amssymb}

\begin{document}

\title{Structural phase transitions in multipole traps}

\author{M. Marciante, C. Champenois, A. Calisti, M. Knoop}

\affiliation{Physique des Interactions Ioniques et Mol\'eculaires (UMR
6633), CNRS et Aix-Marseille Universit\'e,
 Centre de Saint J\'er\^ome, Case C21,
13397 Marseille Cedex 20, France}%
\email{mathieu.marciante@etu.univ-provence.fr}

\date{\today}

\begin{abstract}
A small number of laser-cooled ions trapped in a linear
radiofrequency multipole trap forms a hollow tube structure. We have
studied, by means of molecular dynamics simulations, the structural
transition from a double ring to a single ring of ions. We show that
the single-ring configuration has the advantage to inhibit the
thermal transfer from the rf-excited radial components of the motion
to the axial component, allowing to reach the Doppler limit
temperature along the direction of the trap axis. Once cooled in
this particular configuration, the ions experience an angular
dependency of the confinement if the local adiabaticity parameter
exceeds the empirical limit. Bunching of the ion structures can then be observed and an analytic expression is proposed to
take into account for this behaviour.
\end{abstract}


\maketitle

\section{Introduction}
\label{sec:introduction}

In the wide range of radiofrequency (rf) traps, quadrupole traps
take a particular place due to the parabolic rf potential they create and to the harmonic pseudo-potential it generates. In contrast, higher order traps show a more "rectangle"  shape of the rf electric field which induces a flat pseudo-potential
well with a reduced rf-driven micromotion in the center of the trap
\cite{gerlich92}. This feature finds its application in the spectroscopy  of cold molecular samples \cite{wester09} or in the interrogation of the hyperfine transition of a large atomic sample for frequency
metrology \cite{prestage07}, the ions being  cooled by collisions with a buffer gas.

Laser-cooling of ions trapped in multipole traps has only been performed very recently \cite{okada07} and
experimental observations have demonstrated a tube-like structure for the ion organization \cite{okada09}. This
self-organization in empty core structure results from the competition between Coulomb repulsion and a confining
force very soft in the center of the trap and stiffer towards the walls. For large sample, this behavior is confirmed by the
density profile deduced in the cold charged fluid limit \cite{champenois09} and for smaller ones,  by
simulations of the equilibrium structures \cite{okada07,calvo09,yurtsever11}. For a large number of ions
trapped in a linear multipole trap, the general structure of a crystallized sample is a set of concentric tubes for which the revolution
axis is the longitudinal trap axis. The evolution of the structure with the number of ions and the trapping
parameters is extensively studied in \cite{yurtsever11}. 

\begin{figure}[h]
    \begin{tabular}{cc}
    \includegraphics[width=4.cm,trim=1.5cm 0.cm 2.4cm 0.7cm,clip]{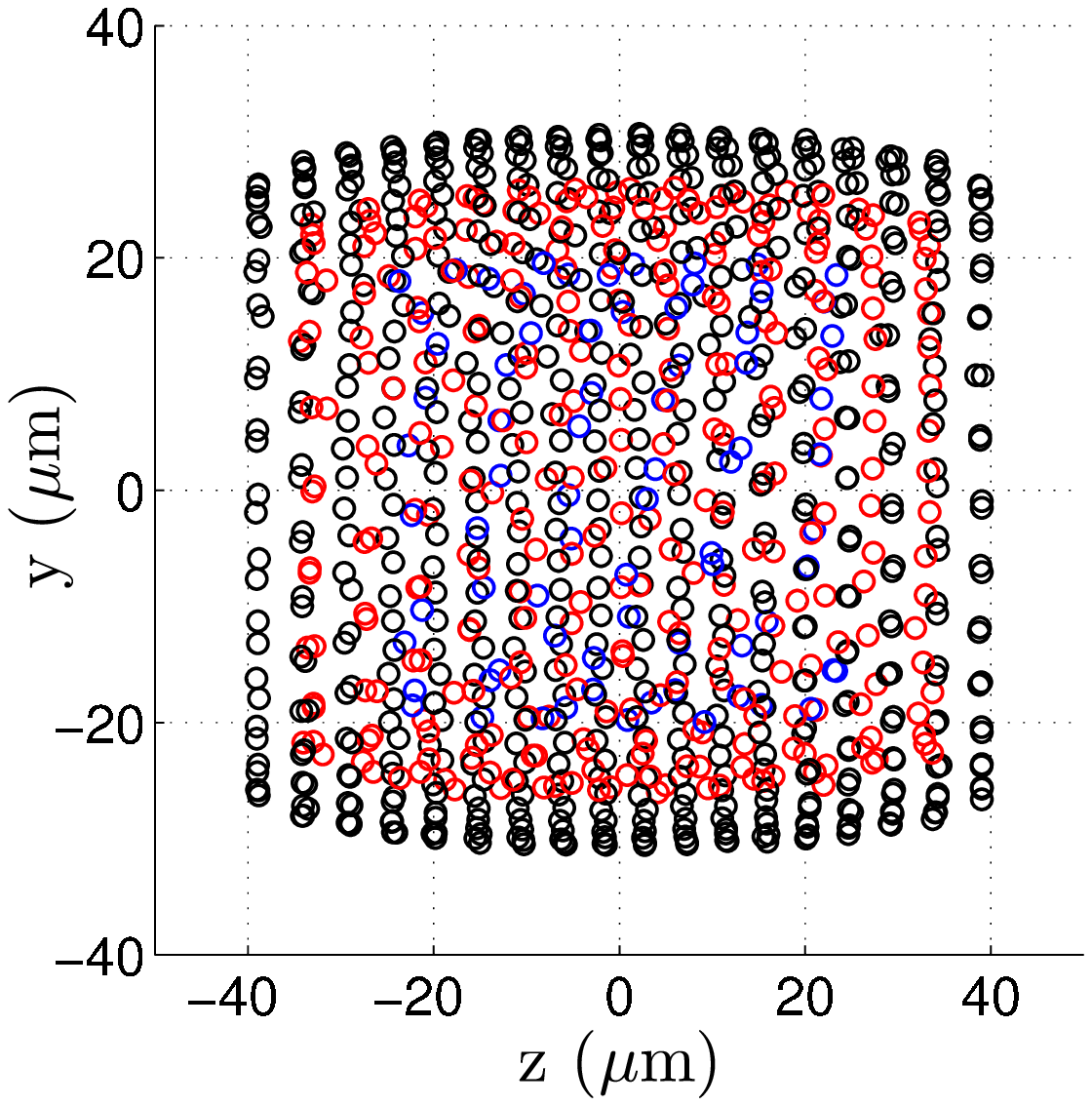} &
    \includegraphics[width=4.cm,trim=1.5cm 0.cm 2.4cm 0.7cm,clip]{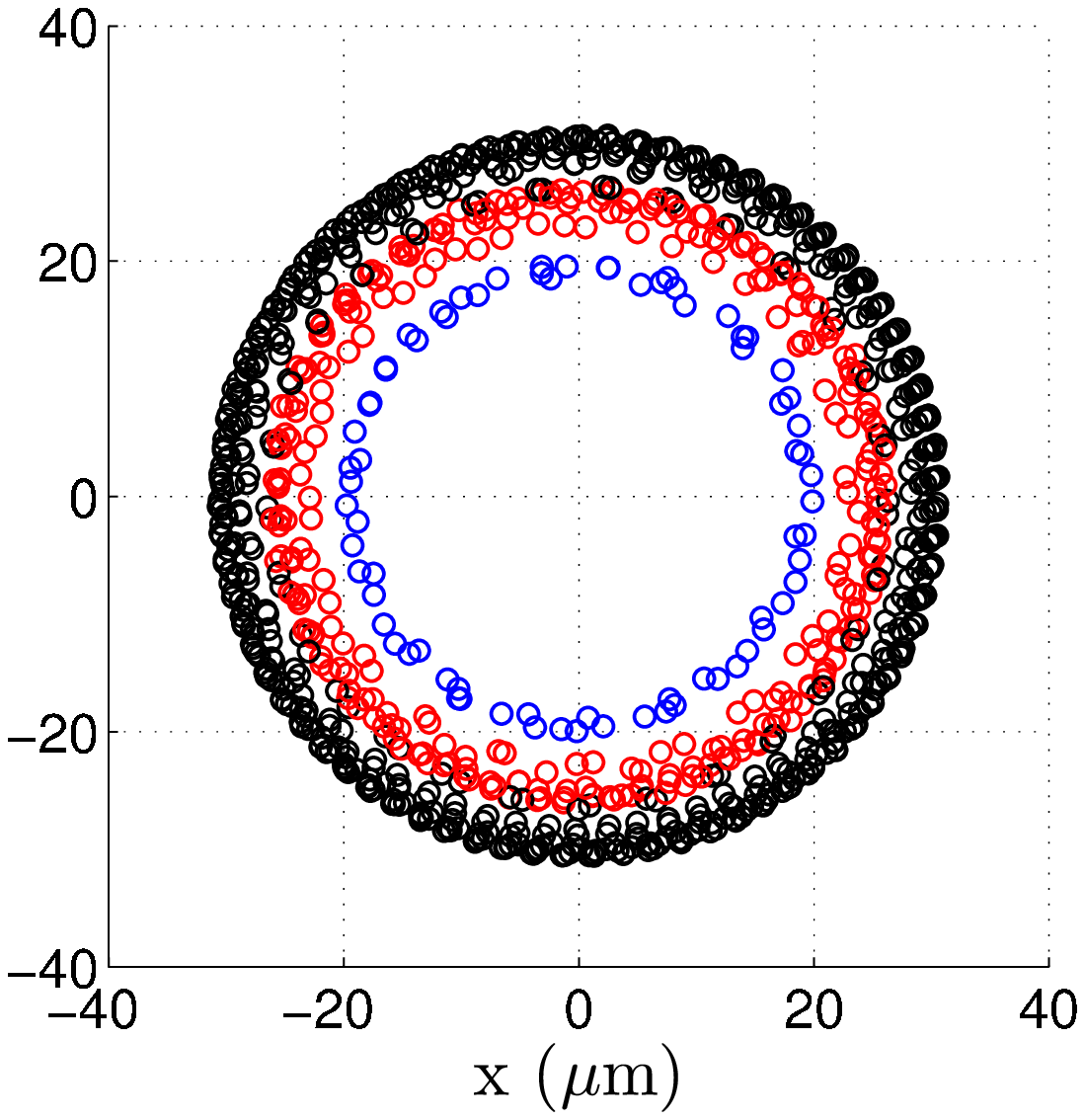}
    \end{tabular}
    \caption{(color online) Structure of a set of 1000 laser-cooled ions in the pseudo-potential of a linear
    octupole trap. The structure is composed of three concentric tubes extending along the trap axis, which is
    in the $z$ direction. The temperature is 0.5~mK in the three directions of space.}
    \label{fig_empty_core}
\end{figure}

As an example, figure~\ref{fig_empty_core} shows the
simulated structure of 1000 cold calcium ions inside the pseudo-potential of a linear octupole trap,
laser-cooled to the Doppler limit temperature of 0.5~mK.
For the same trapping parameters and a smaller number of ions, the structure reduces to a single tube, formed
of parallel rings of ions aligned along the trap axis. For a decreasing number of ions, the tube reduces to
two rings then to a single ring. Ring- or tube-like structures offer the possibility to study dynamics 
in a two dimensional configuration. Furthermore, ion rings have symmetry properties that can be exploited in
high-resolution spectroscopy. Among these highly symmetric structures, the single ring configuration has the
outstanding property to be a true two dimensional system where the motional temperature  along the symmetry axis can reach the
Doppler limit when the ions are laser-cooled, despite the rf-driven motion in the transversal plane
\cite{champenois10}. A previous numerical study \cite{marciante10} of the cooling properties of ion structures
formed in quadrupole traps has shown that for structures of reduced spatial dimensions, Coulomb interaction does
not always couple the different directions of motion of the ions. This decoupling inhibits the thermal transfer
between the decoupled directions of the motion and allows to reach the Lamb-Dicke regime for the ion motion
along the longitudinal trap axis \cite{champenois10}.

In the present article we use molecular dynamics simulation of laser-cooled ions in a linear octupole trap to study their
structural phase transition from a two-ring to a one-ring configuration.
The model used to simulate the ion
dynamics is described in section~\ref{sec:framework} together with
some brief reminders on linear multipole traps.
Section~\ref{sec:transition} presents an analysis of the structural transition itself and enlightens the role of the rf-driven motion  on this transition. In section~\ref{sec:above}, we extend our study by increasing the adiabaticity parameter $\eta_{ad}\left(r\right)$ beyond the empirical limit.

\section{Framework of the model}
\label{sec:framework}

The numerical simulations use molecular dynamics, based on the symplectic velocity-Verlet algorithm to
integrate Newton's equations. We have simulated the dynamics of a set of 100 ions in the potential of a linear
octupole trap, cooled by Doppler laser-cooling. Ions are thus subject to conservatives forces deriving from the
trap electric potential and the Coulomb inter-ion potential, and dissipative forces due to Doppler cooling. The
general form for the ideal rf potential of a linear $2k$-pole trap can be written in polar coordinates like
\beq
    \Phi_{2k}\left(r,\varphi,t\right) =
    q_e V_{2k} \left(\frac{r}{r_0}\right)^{k} \cos\left(k\varphi\right)\ \cos\left(\Omega t\right)\ ,
    \label{eq_phirf}
\eeq
where $V_{2k}$ is the amplitude of the rf potential applied to the $2k$ rods, oscillating at frequency
$\Omega/2\pi$, $r_0$ is the inner radius of the trap, and $q_e$ is the electric charge of the ions. "Unless otherwise specified",
 we use the potential of an octupole trap (2$k = 8$) to run the simulations. This rf potential
induces an rf-driven motion which produces a ponderomotive force, which leads to the radial confinement of the
ions. It can be derived from the static pseudo-potential \cite{gerlich92}:
\beq
    \Psi_{2k}\left(r\right) =
    \frac{k^2}{4}\ \frac{q_e^2V_{2k}^2}{m r_0^2 \Omega^2} \left(\frac{r}{r_0}\right)^{2k-2}.
    \label{eq_pseudo}
\eeq
The  motion in this pseudo-potential has a time scale larger than an rf period and is called macro-motion.

The axial confinement is achieved by a harmonic static potential well characterized by the frequency
$\omega_z/2\pi$, which scales like $\sqrt{q_e/m}$. Obeying Laplace's law, this static potential has a deconfining
contribution in the radial plane. Hence, for each ion of mass $m$ and position $\left(r,\varphi,z\right)$, the
total trapping potential $U_\mathtt{trp}$ writes:
\beq
    U_\mathtt{trp}\left(r,\varphi,z,t\right) =
    \Phi_{2k}\left(r,\varphi,t\right) +\frac{1}{2}\ m\ \omega_z^2\ \left(z^2-\frac{r^2}{2}\right)\ .
\eeq
When the order of the multipole is higher than $2k = 4$, the deconfining contribution of the axial confinement
leads to a shift of the minimum of the trapping potential in the radial plane, from the centre to a circle of
radius $R_\mathtt{min}$:
\beq
    R_\mathtt{min}=r_0
   \left( \frac{1}{k^2\left(k-1\right)}\ \frac{m r_0^2 \Omega^2}{q_e^2 V_{2k}^2}\ m r_0^2 \omega_z^2\ \right)^{1/(2k-4)} .
    \label{eq_rmin}
\eeq
Hence, contrary to a quadrupole trap configuration, the local minimum of the trapping potential is no more
superposed to a node of the rf electric field \cite{marciante11}. Throughout this article, the confinement
parameters and number of ions are such that the ions sit very close to the potential minimum and they do
undergo micro-motion.

The Coulomb interaction is taken into account by the total pair interaction potential, which writes:
\beq
    U_\mathtt{C} = \frac{q_e^2}{4\pi\varepsilon_0}\
                \sum_{i = 1}^{N_p-1} \sum_{j = i+1}^{N_p} \frac{1}{|\delta_{ij}|}\ ,
\eeq
where $N_p$ is the number of ions and $\delta_{ij}$ is the relative distance between ions $i$ and $j$. The
Doppler laser-cooling is modeled using velocity kicks simulating the ion's recoil, which happens depending on
the probability laws of absorption (given by the stationary solution of the optical Bloch equations) and
emission (given by the exponential decay law). For symmetry reasons, laser-cooling is applied in the three
directions of the trap. Details about the model we used for the cooling are given in \cite{marciante10}.

Initial conditions are prepared by giving to each ion an initial position randomly chosen in the inner 10\% of
the radius of the trap and a random velocity obeying a thermal distribution corresponding to 1~K. Ions are
trapped by the rf electric field and are subject to Doppler cooling. The system is ready for
further simulations when the temperatures are stationary. The temperature is one of the relevant parameters to
get insight into ion dynamics. As the rf-driven motion is not a thermal motion, it has to be subtracted
from the velocities before computing the temperature \cite{prestage91b,schiffer00}. In practice, this is done by averaging the velocity of the ions over an rf period. Each averaged
component of the velocity is then taken into account to define the temperature associated to each direction of
the motion. Explicit details about the computation of the temperatures can be found in \cite{marciante10}.

All the forthcoming  results assume  perfect trap operation and neglect defects of the potential due to patch potential and/or misalignments of the electrodes as well as collisions with the residual gas.  Patch potentials with a highly non-uniform spatial repartition could destroy the highly symmetric structure formed by few ions in a linear multipole trap. For high precision spectroscopy in  multipole trap, more extra electrodes than for a line in quadrupole trap may be required to compensate for patch potentials and their gradient. Misalignment of the electrodes has been  identified in a 22-pole trap \cite{otto09}  as responsible for 10 extra potential minima, which cause bunching of the ions in these local potential wells.  As for collisions, they may perturb the formation process of the crystal but the trapping conditions chosen for these studies imply a strong confinement in the axial and in the radial direction, of the same order of magnitude as the one met in quadrupole trap. Therefore, the stationary structures described in this article should survive to collisions like ion chains do. This strong confinement may seem unrealistic regarding the smooth profile of the trapping field in the center of the trap. The scaling law for the rf electric field in an octupole is $V_8 / {r_0}^4$, and to reach a $R_{min}$ value of $120 \mu$m with $ \omega_z=1$ MHz and $\Omega=20$ MHz like chosen in the following, it  takes an amplitude $V_8$ equal to 208 V in a  0.5 mm inner radius trap, which is very accessible. As for quadrupole trap dedicated to single ion experiment, a multipole trap  able to keep a single ring stable has to be specifically designed to that purpose.

\section{Double-ring to single-ring structural transition}
\label{sec:transition}

\begin{figure}[h]
    \begin{tabular}{cc}
    \includegraphics[width=4.2cm,trim=6.6cm 0.cm 7.4cm 0.7cm,clip]{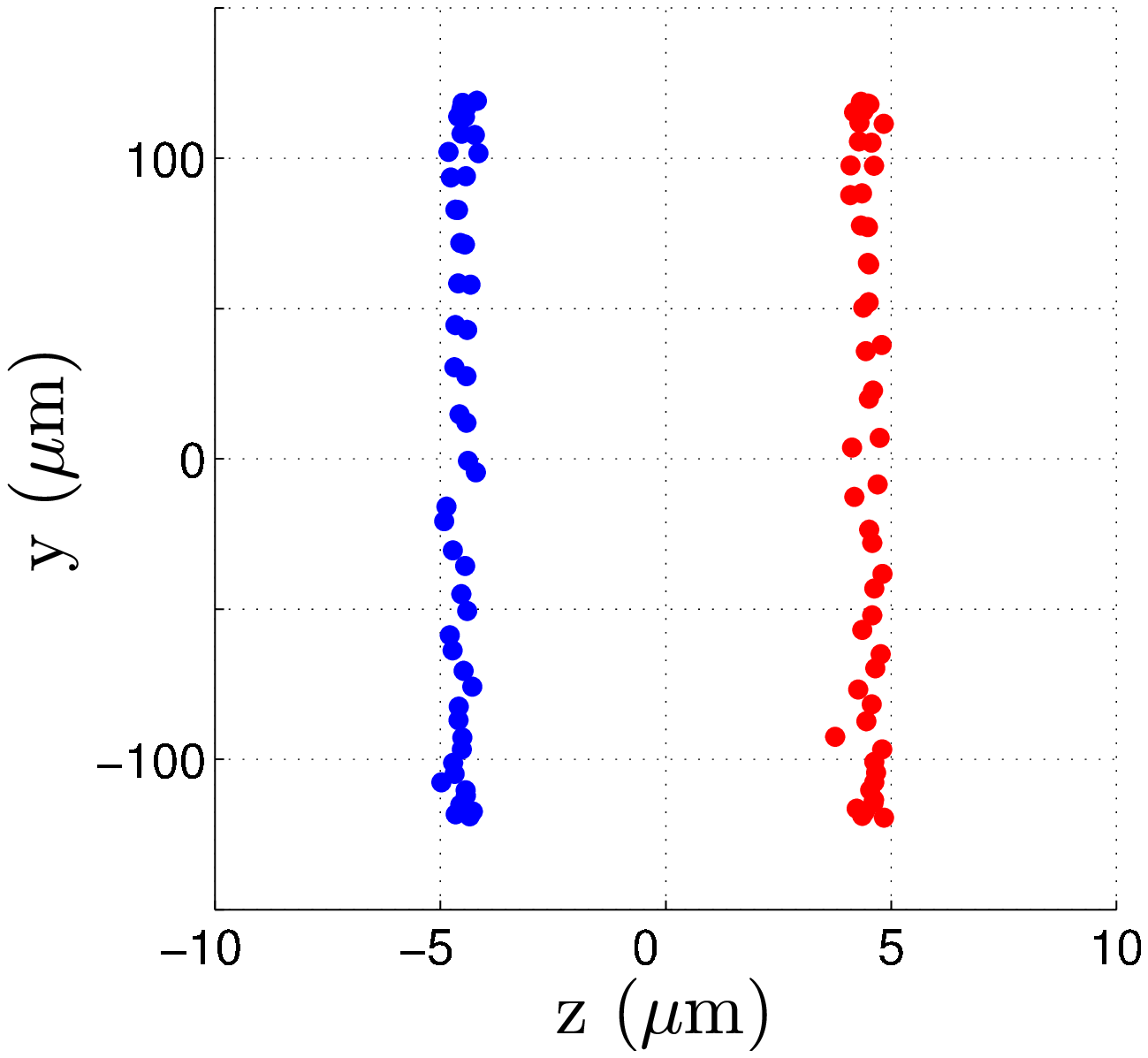} &
    \includegraphics[width=4.cm,trim=6.4cm 0.cm 7.4cm 0.7cm,clip]{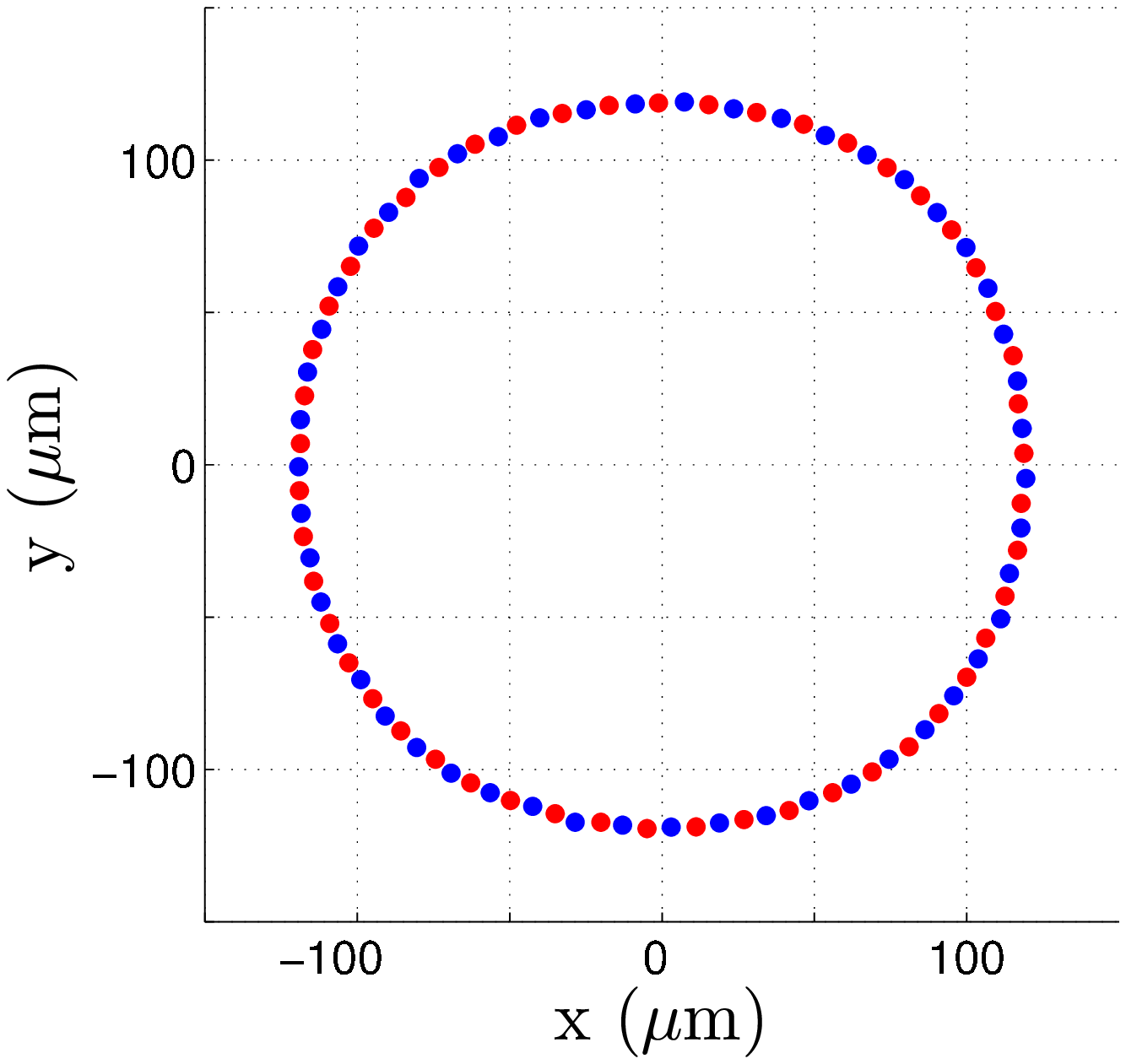}
    \end{tabular}
    \caption{(color online) Double-ring structure of 100 laser cooled ions, in the rf potential of an octupole
    trap. The axial confinement frequency is $\omega_z/2\pi = 0.5$~MHz, leading to an axial separation between
    the two rings of about 9~$\mu$m. The picture corresponds to the instant of the maximal amplitude of the micro-motion.}
    \label{double-ring}
\end{figure}

\begin{figure}[h]
    \includegraphics[width=7.5cm,trim=1.5cm 0.0cm 2.5cm 1.5cm,clip]{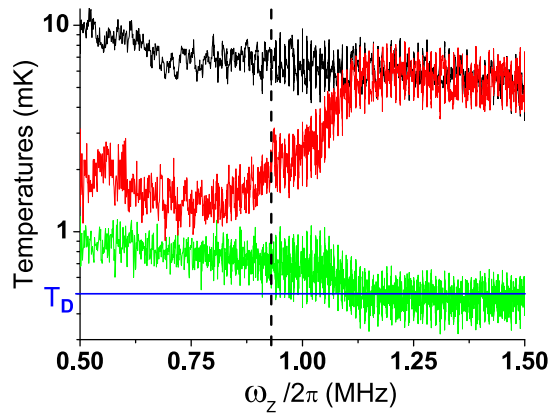}
    \caption{(color online) Temperatures associated to the motion in a polar coordinate system $T_r$ (black upper curve),
    $T_\varphi$ (red medium curve) and $T_z$ (green bottom curve), when the axial confinement frequency
    $\omega_z$ is increased. The vertical dashed line marks the theoretical transition frequency from the
    double-ring to the single-ring structure. The horizontal blue line materializes the Doppler
    limit temperature.}
    \label{fig_temperatures}
\end{figure}

The trapping parameters are such that the Coulomb repulsion is negligible compared to the deconfining effect
associated to the axial confinement and the radial position of the ions is controlled by the potential minimum
$R_\mathtt{min}$. Depending on the number of ions $N_p$ and the applied axial potential, the stationary
structure is a single or a double ring. By stability consideration, one can show  that there
is a critical radius $R_\mathtt{crit}$ below which the equilibrium configuration is a double ring \cite{champenois09}. This
critical radius only depends on $N_p$ and $\omega_z$ through
\beq
    R_\mathtt{crit} \approx \left(\frac{q_e^2/4\pi\varepsilon_0}{2\ m\ \omega_z^2}\right)^{1/3}\ \frac{N_p}{\pi}\ ,                 \label{eq_Rlim}
\eeq
For a set of 100 ions and $\omega_z/2\pi = 0.5$~MHz, the stability critical radius $R_\mathtt{crit}$ is
$181~\mu$m. The rf electric potentiel is chosen such that the radius $R_\mathtt{min}$ is $120~\mu$m and the
ions organize themselves in two rings of 50 ions separated by an axial distance of about $9~\mu$m. This
double-ring structure can be seen on figure~\ref{double-ring}, for an rf phase corresponding to the maximal
amplitude of the micro-motion.
The pattern formed by the two ion rings is equivalent to an annular zig-zag structure \cite{waki92}, the positions of two consecutive ions being on different sides of the $z = 0$ plan. This
perfect annular zig-zag is obtained by choosing an even number of ions. In this configuration, the temperature
associated with the axial component of the motion is $T_z \approx 1$~mK (see figure \ref{fig_temperatures}). If
the velocity of the ions is expressed in cartesian coordinates, the $T_x$ and $T_y$ temperatures are identical
and are approximately equal to 12~mK. The difference between these values is attributed to a partial
thermalization of the degrees of motion, the higher temperature being induced by rf-driven motion in the radial
plane. This motion induces rf-heating by which the kinetic energy of the rf-driven motion is transferred to the
macro-motion \cite{ryjkov05}. It also reduces the Doppler cooling efficiency by the Doppler shift on the cooling
transition. The temperature discrepancy shows that the Coulomb coupling in the two-ring structure is not strong
enough to thermalize the motion in the radial plane with the motion along the axis \cite{marciante10}. Inside
the radial plane itself, the use of cylindrical coordinates reveals an important discrepancy in the
temperature associated to the radial motions ($T_r \simeq 10$~mK) and the one associated to the tangential
motion ($T_\varphi \simeq 2$~mK, see figure \ref{fig_temperatures}).

Starting from this configuration, we study the double to single ring transition. To that purpose, the
characteristic frequency of the axial confinement $\omega_z/2\pi$, is slowly increased from 0.5~MHz to 1.5~MHz.
At the same time, the amplitude of the rf electric field is raised to balance the enhancement of the static deconfining effect in
the radial plane, keeping the minimum of the trapping potential, $R_\mathtt{min}$, and thus the radius of the
ring, constant (see equation~\ref{eq_rmin}). For a constant number of ions, the structural transition
is expected from equation \ref{eq_Rlim} to take place for $\omega_z/2\pi=0.93$~MHz. It is verified that the
variation of the trapping parameters is sufficiently slow to consider the system to be in a stationary state during
the simulation. Particular cares are taken for values above 0.92~MHz for which the increase of the confinement
parameters is halted every 0.1~MHz, by time periods of 10~ms.

\begin{figure}[h]
    \includegraphics[width=7.5cm,trim=2.cm 0.0cm 0.5cm 1.5cm,clip]{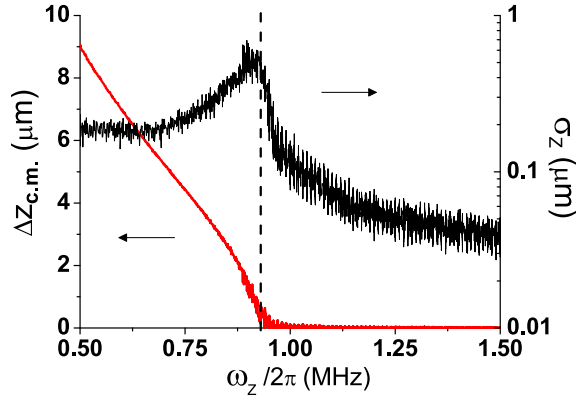} \\
    \caption{(color online) Relative distance between the c.m. of the two rings, $\Delta z_\mathtt{cm}$ (red curve,
    left axis), and standard deviation of the axial component of the ion's positions from the c.m. position of the
    ring of which they belongs, $\sigma_z$ (black curve, right axis), function of the increasing axial confinement
    frequency $\omega_z$. The vertical dashed line marks the theoretical transition from the double-ring to
    single-ring configuration.}
    \label{fig_cm}
\end{figure}

\begin{figure}[h]
    \begin{tabular}{cc}
    \includegraphics[height=4.5cm,trim=6.4cm 0.cm 7.4cm 0.7cm,clip]{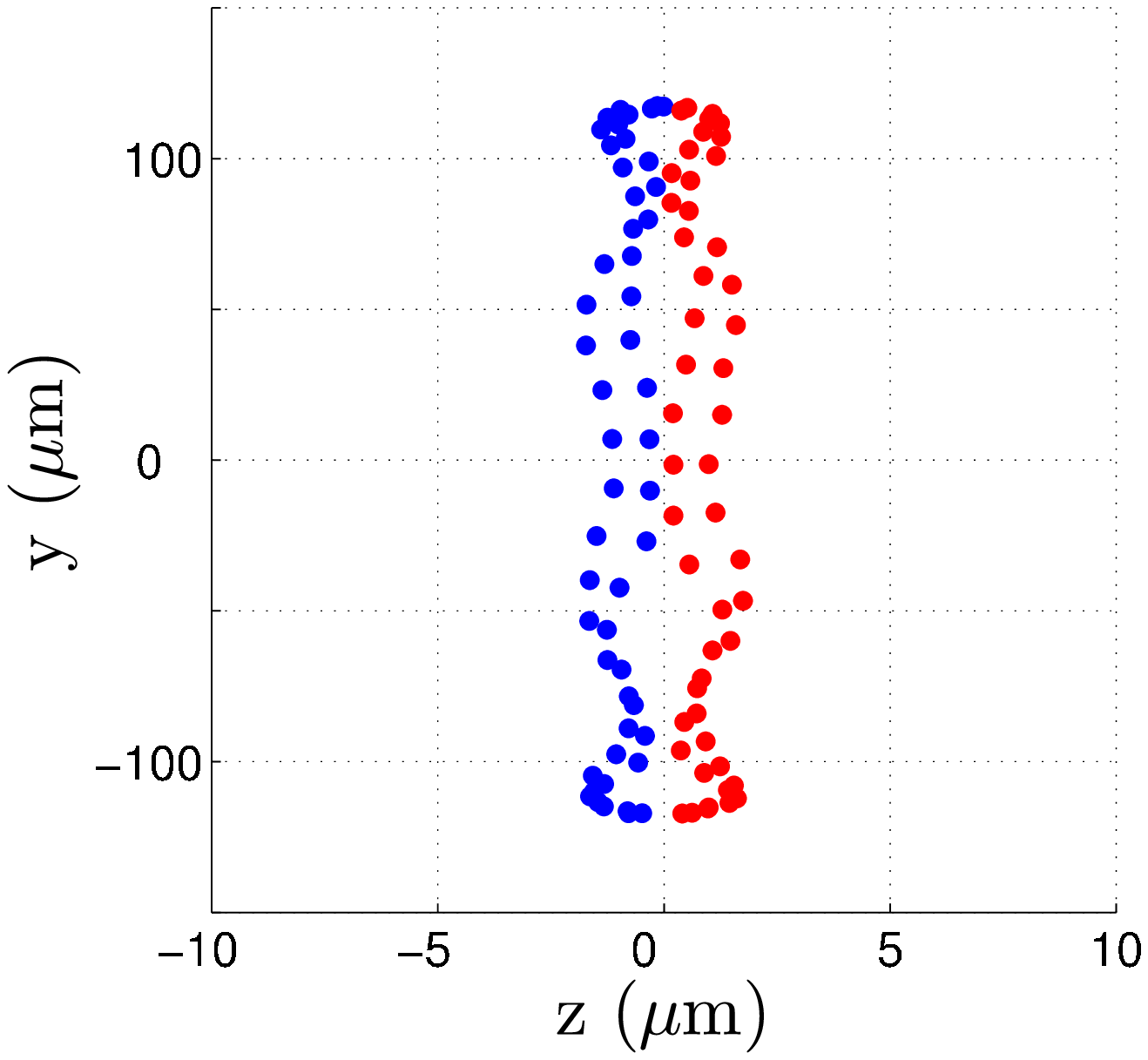} &
    \includegraphics[height=4.5cm,trim=9.5cm -0.3cm 9.cm 1.0cm,clip]{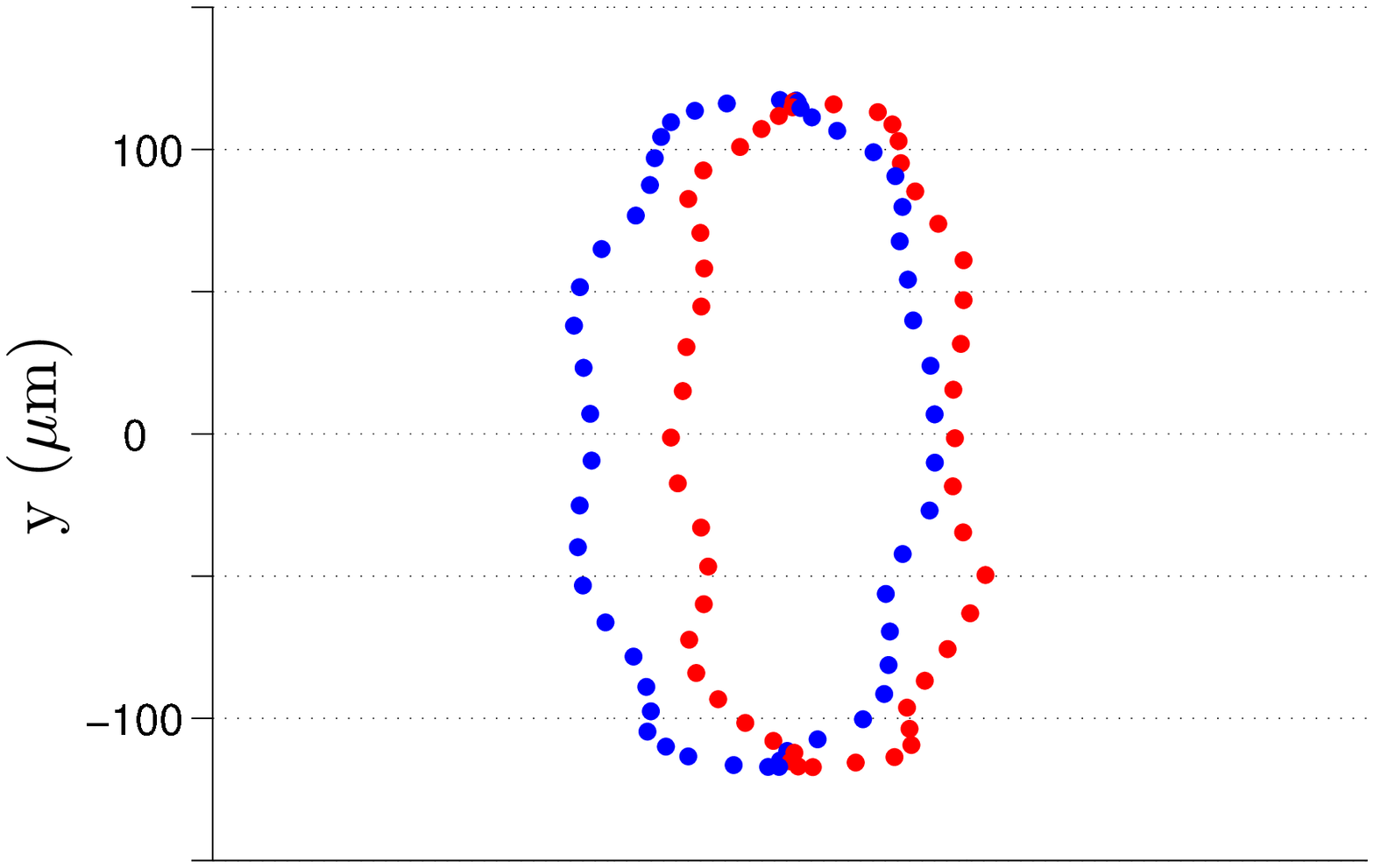}
    \end{tabular}
    \caption{(color online) Distorsion of the double-ring structure viewed in the $\left(z,y\right)$-plane (left
    picture), and from a point of view rotated of $-\pi/4$ around the $y$-axis. The structure is stationary over
    the rf-period time scale. This picture is taken when the distance between the two centers of mass is about
    2~$\mu$m. }
    \label{fig_deformation}
\end{figure}

The structural transition is first monitored by the distance between the center of mass (c.m.) of the two
rings, $\Delta z_\mathtt{cm}$, given on the left axis of figure~\ref{fig_cm}, as a function of the
characteristic frequency $\omega_z/2\pi$. The value for which the two rings merge is in good agreement with
the theoretical frequency of 0.93~MHz, for which the structural transition is predicted.
 \\
The standard deviation of the axial component of the ions' positions from the c.m. position is also plotted on
figure~\ref{fig_cm}. For $\omega_z/2\pi = 0.5$~MHz, the standard deviation $\sigma_z$ is due to the axial
thermal motion around the equilibrium position. Increasing $\omega_z$, the axial temperature (shown on
figure~\ref{fig_temperatures}) drops while the local steepness of the potential in the axial direction is
increased. Hence, the axial dispersion on ions' positions which is due to thermal motion is expected to decrease
when $\omega_z$ grows. Figure \ref{fig_cm} shows that it becomes maximum around the structural transition before
decreasing to a lower value than the initial one. To understand the transition process, we focus on the ion
organization just before the structural transition. While the two rings are getting closer, a distortion of the
ring structures progressively appears, visible on figure~\ref{fig_deformation} and stationary over the rf
period. The distortion does not affect the equilibrium radius of the structures but the axial inter-ion distance
is smaller for ions localized close to the rf electrodes than for those which are trapped far from the electrodes (with the convention defined by
equation~\ref{eq_phirf}, electrodes are at angular positions $\varphi_n = n\pi/4$). This behaviour is attributed
to micro-motion whose amplitude $\vec\varepsilon_{\mu}\left(r,\varphi,t\right)$ has the analytic form:
\beq
    \vec\varepsilon_{\mu}\left(r,\varphi,t\right) =  \frac{q_e  V_8r^3}{m r_0^4\ \Omega^2}\
          \left(\cos\left(4\varphi\right) \hat r -\sin\left(4\varphi\right) \hat\varphi\right)\
          \cos\left(\Omega t\right) ,
    \label{eq_micromvt}
\eeq
When ions are localized in front of an rf-electrode, the micro-motion amplitude is almost
exclusively radial and has no significant consequence on the relative distance of two neighbouring ions
belonging to different rings. On the contrary, when ions are localized at an inter-electrode angular position,
the micro-motion is mostly tangential to the ring structure and depends like $\sin(4\varphi)$ on the ion
angular position. Two neighbouring ions, separated by an angle $2\pi R/N_p$, do not have the same tangential
amplitude of micro-motion, producing an oscillation of their angular separation. Since the neighbouring ions do
not belong to the same ring, this variation of angular positions brings them face to face, increasing their
Coulomb repulsion in the axial direction. The comparison with simulations run in the pseudo-potential
(equation~\ref{eq_pseudo}) confirms that this behaviour is induced by the rf motion.
 \\

The evolution of the distortion during the increase of $\omega_z$ explains the increase of standard deviation
$\sigma_z$ roughly situated between 0.75~MHz and 0.93~MHz (Figure \ref{fig_cm}). When the transition frequency is reached, the c.m. of
the two structures match roughly, but the standard deviation takes its maximal value, showing that the structure
is not yet a planar ring. For higher values of $\omega_z$, the standard deviation shows a steep downward slope,
until a frequency value slightly lower than 1~MHz. The structure can then be considered as a planar single-ring
structure.

The distortion of the structure described above and the coupling between the tangential and axial component of
motion it induces, explain the initial discrepancy between the radial and tangential temperatures when the
structure is a double-ring (see figure~\ref{fig_temperatures}). The axial motion being not subjected to
micro-motion, it can reach lower temperature than the motions in the radial plane. The coupling
between axial and tangential components of the motion allows to dissipate a part of the tangential thermal
kinetic energy through the axial cooling. As the frequency $\omega_z$ is increased, the temperature $T_\varphi$
increases progressively and meets the temperature associated to the radial directions, $T_r$. At the same time,
the axial temperature $T_z$ shows a steeper drop between 1~MHz and 1.25~MHz, reaching the Doppler limit
temperature at the frequency where the two other temperatures join. This behaviour suggests that the axial
component of the motion of the ions is then decoupled from the two transverse directions, like expected for a
full two dimensional system \cite{marciante10}.

The comparison between Figures \ref{fig_cm} and \ref{fig_temperatures} shows that for this small 100-ion system, informations on the dimensionality of the system given by the variation of the ions' positions, $\sigma_z$, and the temperatures of the system do not coincide. The decoupling observed on the temperature arises when the $\sigma_z$ parameter already shows a planar structure. As a consequence, each of these parameters may be a relevant criteria to determine the structural phase transition.

\section{Above the adiabatic limit}
\label{sec:above}


The single ring structure presented in the preceding section seems very promising for applications such high
precision spectroscopy, quantum information or for metrological applications \cite{champenois10,olmos09}. For
all of these, it is mandatory to reduce the amplitude of motion of the ions to below the chosen transition
wavelength (the Lamb-Dicke regime \cite{wineland78}) to cancel the first order Doppler effect. For
trapped ions, Doppler laser-cooling is sufficient if the trapping frequency along the laser propagation
direction is large enough. In practice, for calcium ions cooled to the Doppler limit, a motional  frequency of
1~MHz allows the ions to reach the Lamb-Dicke regime for an optical transition. One of the main drawbacks of the
ring structure compared to its linear counterpart, the ion chain, found in quadrupole traps is the non null rf-driven motion
(micro-motion) at their equilibrium positions. Figure \ref{fig_temperatures} shows that in the single ring
configuration, the motion along the axis is decoupled from the motion in the radial plane where micro-motion,
and thus rf-heating, take place. The Lamb-Dicke regime can thus be reached for laser propagating along the
trap axis.

\begin{figure}[h]
  \begin{tabular}{ccc}
    \includegraphics[width=3.3cm,trim=0.3cm 0.5cm 0.5cm 0.5cm,clip]{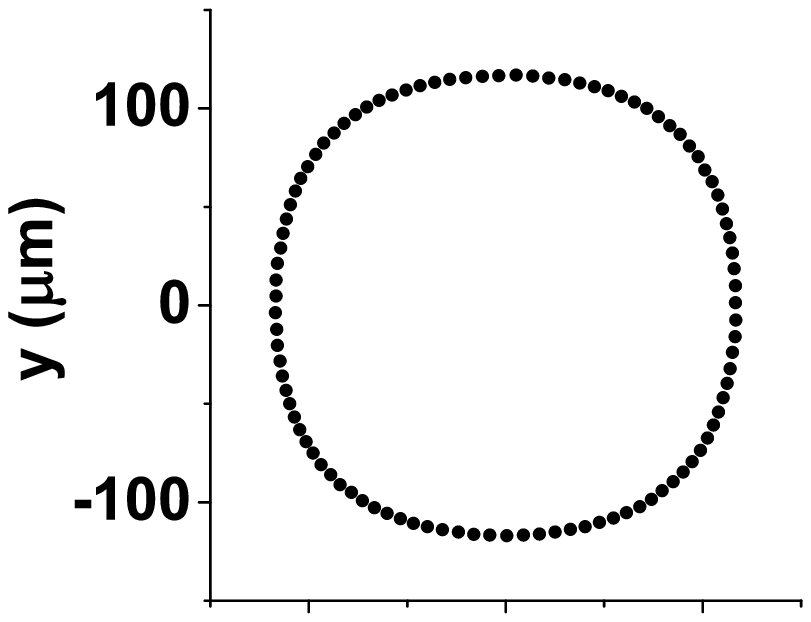} &
    \includegraphics[width=2.3cm,trim=1.0cm 0.5cm 0.5cm 0.5cm,clip]{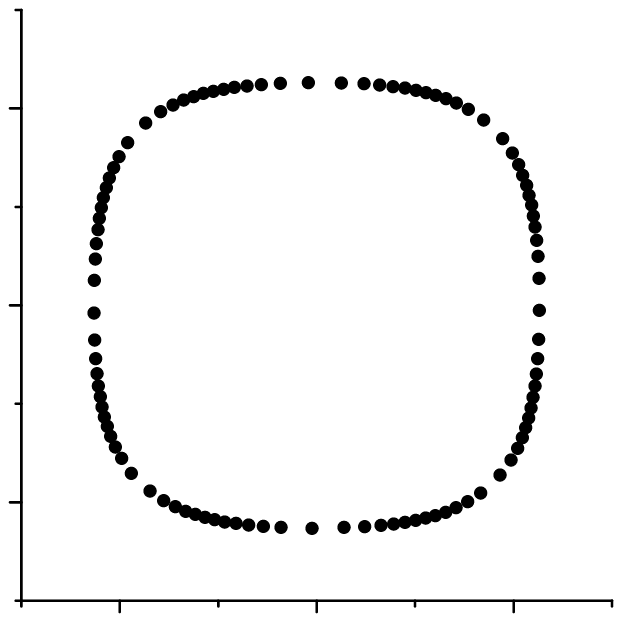} &
    \includegraphics[width=2.3cm,trim=1.0cm 0.5cm 0.5cm 0.5cm,clip]{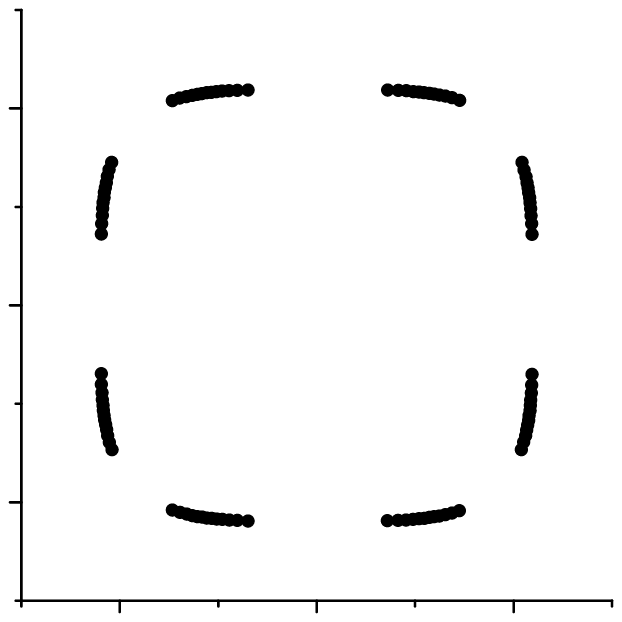}    \\
    \includegraphics[width=3.3cm,trim=0.5cm 0.5cm 0.5cm 0.5cm,clip]{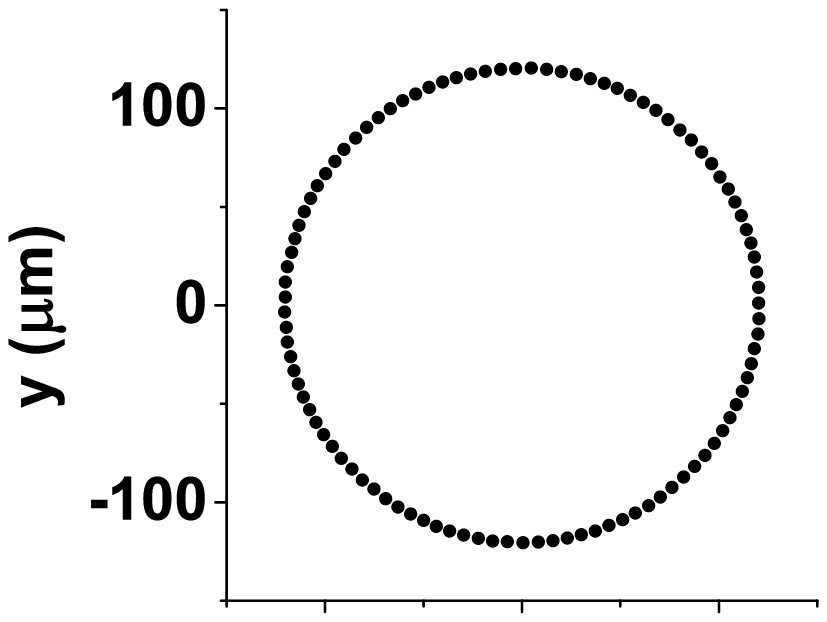} &
    \includegraphics[width=2.3cm,trim=1.0cm 0.5cm 0.5cm 0.5cm,clip]{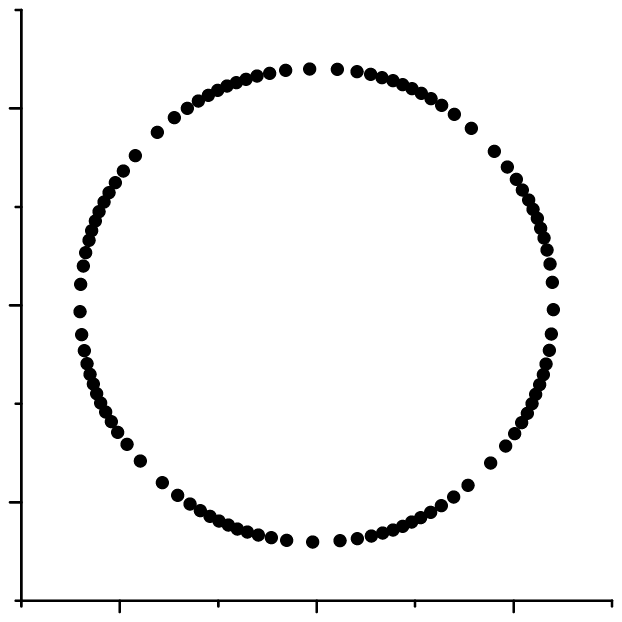} &
    \includegraphics[width=2.3cm,trim=1.0cm 0.5cm 0.5cm 0.5cm,clip]{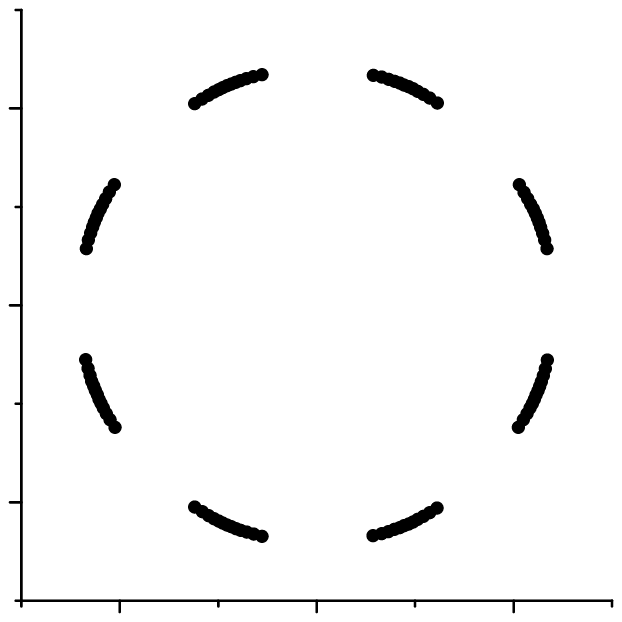}    \\
    \includegraphics[width=3.3cm,trim=0.5cm 0.5cm 0.5cm 0.5cm,clip]{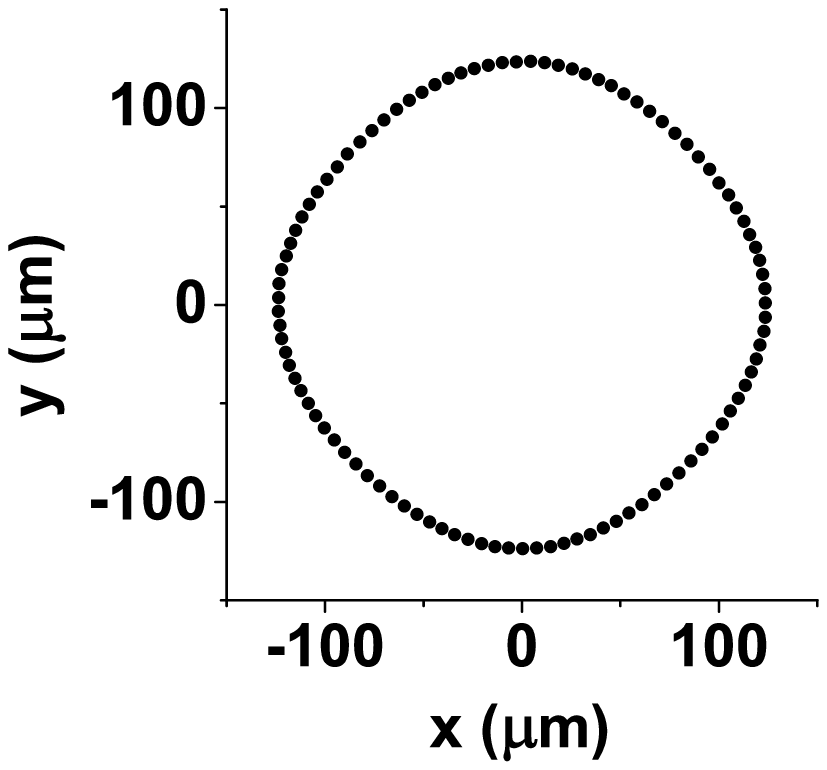} &
    \includegraphics[width=2.3cm,trim=1.0cm 0.5cm 0.5cm 0.5cm,clip]{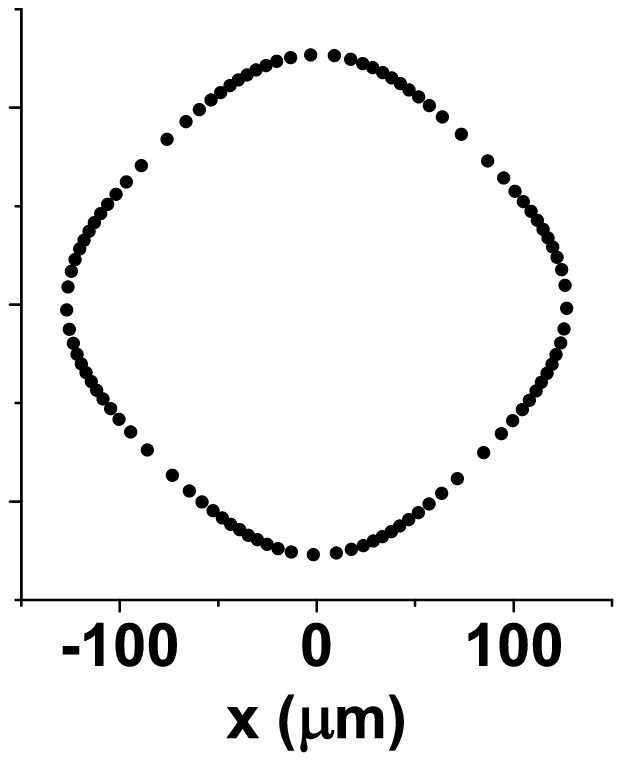} &
    \includegraphics[width=2.3cm,trim=1.0cm 0.5cm 0.5cm 0.5cm,clip]{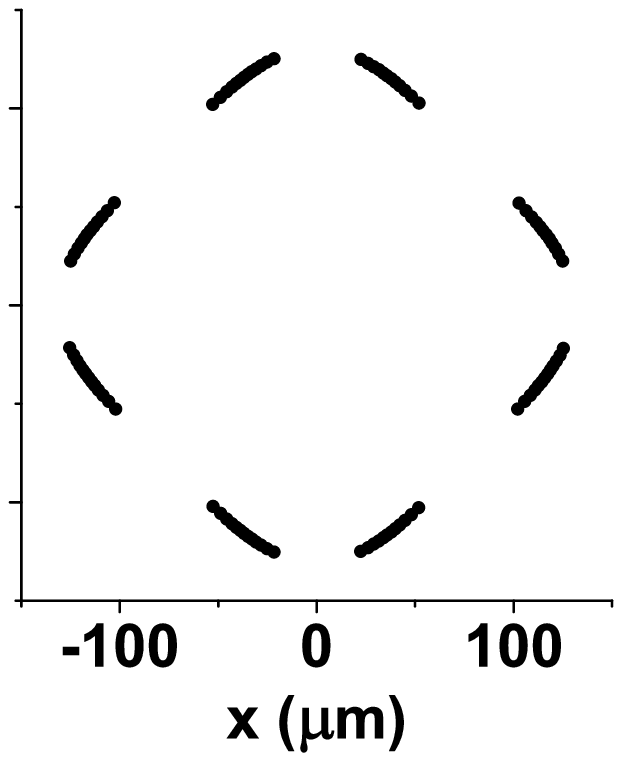}    \\
    \end{tabular}
    \caption{ View in the radial plane of structures obtained for an equilibrium radius $R_\mathtt{min} =
    120~\mu$m and, from left to right, $\omega_z = $ 1~MHz, 2~MHz and 3~MHz. From top to bottom, the pictures correspond to three phase of the rf:   $\cos\left(\Omega t\right) = -1; 0; 1$.}
    \label{fig_substructures}
\end{figure}

To quantify the impact of the micro-motion on the ion dynamics and laser cooling, the Mathieu parameter $q_x$,
relevant for this issue in quadrupole traps, can be extrapolated to multipole trap and is then related to the
more general concept of adiabaticity of the trajectories \cite{gerlich92}. The adiabaticity parameter
$\eta_{ad}$ depends on the radial position of the ions inside the trap and at the potential minimum
$R_\mathtt{min}$, it takes the simple expression \cite{champenois09}:
 \beq
    \eta_{ad}\left(R_\mathtt{min}\right) = 2\ \sqrt{k-1}\ \omega_z/\Omega\ .
    \label{eq_eta}
\eeq
It has been demonstrated, numerically and experimentally in a 22-pole \cite{mikosch07,mikosch08,wester09}, that
for $\eta_{ad}\left(r\right) < 0.36$ the trajectories are stable but few studies concern this adiabaticity
parameter in the context of cold samples. In this section, we focus on the role of the micro-motion on the
self-organization of the ions and relate it to the adiabaticity parameter. Figure \ref{fig_substructures} shows
the single ring structure formed by 100 ions located at the same radial potential minimum and trapped by an
axial potential characterized by $\omega_z/2\pi$ equal to 1~MHz ($\eta_{ad} = 0.173$), 2~MHz ($\eta_{ad} =
0.346$) and 3~MHz ($\eta_{ad} = 0.520$). For increasing $\omega_z$, we observe a bunching of the ions at the
inter-electrode angular positions.

This bunching is not observed in the pseudo-potential approach of the trapping electric field and is attributed
to the micro-motion of the ions. Some insight in the role of micro-motion can be gained by considering a cold
sample sitting close to $R_\mathtt{min}$, where the radial trapping potential can be approximated by a local harmonic
potential. Assuming a null potential for $R_\mathtt{min}$, it can be written like
\beq
        U_\mathtt{loc}\left(R_\mathtt{min}+\delta r\right)
        = \frac{1}{2}\ m\ \omega_\mathtt{loc}^2 \left(\delta r\right)^2\ ,
        \label{eq_uloc}
\eeq
with the characteristic frequency $\omega_\mathtt{loc} = \sqrt{k-2}\ \omega_z$ \cite{champenois09}. The ion
motion is assumed to be due only to the rf-driven micro-motion whose amplitude can be written like:
\beqa
    \vec\varepsilon_{\mu}&&\left(R_\mathtt{min},\varphi,t\right)
    = \\
     &&\|\vec\varepsilon_{\mu}\left(R_\mathtt{min}\right)\|
          \left(\cos\left(4\varphi\right) \hat r -\sin\left(4\varphi\right) \hat\varphi\right)\
          \cos\left(\Omega t\right) \nonumber ,
\eeqa
where the amplitude of the micro-motion $\|\vec\varepsilon_{\mu}\left(R_\mathtt{min}\right)\|$ is easily deduced
from equation~\ref{eq_micromvt}. Only the radial contribution of this motion engenders a significative potential
energy variation during the rf-period. Its time average depends on the angular position $\varphi$ like
\beq
        \overline{U_\mathtt{loc}}\left(R_\mathtt{min},\varphi\right) = \frac{1}{4}\ m\ \omega_\mathtt{loc}^2\
        \|\vec\varepsilon_{\mu}\left(R_\mathtt{min}\right)\|^2\ \cos^2\left(4\varphi\right)\
        \label{eq:Uloc}
\eeq
and could explain the bunching observed on figure \ref{fig_substructures}.
To test this assumption, we have created a configuration where only two ions bunch in one of the local angular
minimum potential localized around $\varphi_0=(2n+1)\pi/8$. To that purpose, we have simulated a set of 9 ions
in the rf-potential of the same octupole trap, with $\omega_z/2\pi$ initially set at 0.5~MHz, and slowly
increased to 4~MHz, keeping $R_\mathtt{min} = 120~\mu$m constant by increasing the rf potential. The final
stable configuration is then made of seven ions and a unique pair of ions, localized  at each inter-electrode
angular positions.

\begin{figure}[h]
    \includegraphics[width=7.cm,trim=0.7cm 0.0cm 0.0cm 0.0cm,clip]{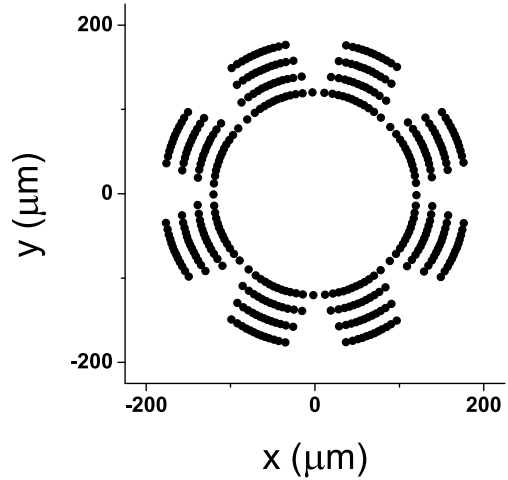}
    \caption{View in the radial plane of a 100-ion structure, for $\omega_z/2\pi = 2~$MHz, and $R_\mathtt{min} =
    $ 120~$\mu$m, 140~$\mu$m, 160~$\mu$m and 180~$\mu$m, varied by modifying the rf-potential amplitude. The
    picture corresponds to a null rf-phase.}
    \label{fig_diffr}
\end{figure}

Assuming a local description $- m \omega_\mathtt{mes}^2 \delta/2$ for the force responsible for the bunching, the measure
of the relative distance $\delta$ of the ions forming the pair allows one to deduce the characteristic frequency
$\omega_\mathtt{mes}$ from the equilibrium condition with the Coulomb repulsion:
\beq
        \omega_\mathtt{mes}= \left(\frac{2\ q_e^2}{4\pi\varepsilon_0 m \delta^3}\right)^{1/2}\ .
\eeq
One can also evaluate the force $\vec F_\varphi$ by deriving the time-averaged micro-motion induced potential
$\overline{U_\mathtt{loc}}\left(R_\mathtt{min},\varphi\right)$, for an angle $\varphi=\varphi_0+\delta\varphi$ :
\beq
\vec F_\varphi\left(\varphi_0+\delta\varphi\right) =- 8 m \omega_\mathtt{loc}^2\
        \frac{\|\vec\varepsilon_{\mu}\left(R_\mathtt{min}\right)\|^2}{R_\mathtt{min}}\ \delta\varphi\ \hat\varphi
\eeq
Writing $\vec F_\varphi$ like $- m \omega_\varphi^2 R_\mathtt{min} \delta\varphi$, one can connect the characteristic frequency
$\omega_\varphi$ to the local radial potential.
The results of numerical simulations show that the predicted bunching frequency $\omega_\varphi$ is larger than the measured frequency
$\omega_\mathtt{mes}$ by a factor $\sqrt{2}$, all along the simulation increasing $\omega_z$ at constant
$R_\mathtt{min}$. The disagreement by a factor $\sqrt{2}$ is also observed when the same simulations are run
with 13 ions in a dodecapole trap. According to the results obtained from the simulations,  the local bunching caracteristic frequency is given by:
\beq
        \omega_\varphi = 2\ \omega_\mathtt{loc}\
        \frac{\|\vec\varepsilon_{\mu}\left(R_\mathtt{min}\right)\|}{R_\mathtt{min}}.
        \label{eq:omega_phi}
\eeq
Eq. \ref{eq:omega_phi} shows how the bunching frequency depends on the
characteristic trapping parameter and ring configuration. Indeed, using the dependence of the micro-motion with
the position in the trap, one can show that $\omega_\varphi$ scales like $\omega_z^2 /\Omega$. Hence, the bunching force scales like $R_\mathtt{min} \omega_z^4 /\Omega^2$. The dependence of this bunching force with $\omega_z$ was illustrated on figure \ref{fig_substructures} for
fixed $R_\mathtt{min}$. Its dependence with $R_\mathtt{min}$ for fixed $\omega_z$ can be seen on figure
\ref{fig_diffr} and confirms an increase of the bunching with $R_{min}$.
These different studies show that the so-called adiabaticity parameter $\eta_{ad}$ is not the relevant criteria
to estimate the role of micro-motion on the ion organization. Indeed, the four structures plotted on figure
\ref{fig_diffr} share the same $\eta_{ad}$. 
This behaviour shows that in the general description of the ion dynamics in a multipole trap, micro-motion has a non-negligible influence not only on the thermal propoerties of the sample but also on its structural properties, and has thus  to be taken into account.

\section{Conclusion}

In this paper, we have numerically studied the double-ring to single-ring structural transition of a Doppler-cooled 100-ion system in the rf potential of an octupole trap. The simulations showed that, in the double-ring configuration, the temperatures associated to each component of the motion expressed in cylindrical coordinates are different. In our initial configuration, the tangential temperature is lower  than the radial one by almost one order of magnitude. This is attributed to a coupling of the tangential motion to the axial one, which is efficiently Doppler-cooled due to the absence of rf-driven motion along the axial direction. When the double-ring structure is compressed, the c.m. of each ring merge at the predicted transition frequency but the structure shows a distorsion which is maximal at this frequency. This behaviour is attributed to the micro-motion of the ions, inducing an axial Coulomb repulsion which can be considered as a pseudo-potential between the ions of the different rings. A planar single-ring structure is thus obtained for a higher value of the predicted transition frequency. When $\omega_z$ is further increased, the axial temperature reaches the Doppler limit value showing a decoupling of the axial and transverse components of the motion.

For a cold ion system, we have also shown that it is possible to increase the adiabaticity parameter beyond the empirical limit, keeping stable ion confinement. At a fixed ring radius, increasing the adiabaticity parameter results in a bunching of the structure, for which an expression of the local bunching frequency is proposed. We have also shown that changing the ring radius at constant value of the adiabaticity parameter leads to the emergence of the bunching phenomenon.


\begin{thebibliography}{0}%
\makeatletter
\providecommand \@ifxundefined [1]{%
 \@ifx{#1\undefined}
}%
\providecommand \@ifnum [1]{%
 \ifnum #1\expandafter \@firstoftwo
 \else \expandafter \@secondoftwo
 \fi
}%
\providecommand \@ifx [1]{%
 \ifx #1\expandafter \@firstoftwo
 \else \expandafter \@secondoftwo
 \fi
}%
\providecommand \natexlab [1]{#1}%
\providecommand \enquote  [1]{``#1''}%
\providecommand \bibnamefont  [1]{#1}%
\providecommand \bibfnamefont [1]{#1}%
\providecommand \citenamefont [1]{#1}%
\providecommand \href@noop [0]{\@secondoftwo}%
\providecommand \href [0]{\begingroup \@sanitize@url \@href}%
\providecommand \@href[1]{\@@startlink{#1}\@@href}%
\providecommand \@@href[1]{\endgroup#1\@@endlink}%
\providecommand \@sanitize@url [0]{\catcode `\\12\catcode `\$12\catcode
  `\&12\catcode `\#12\catcode `\^12\catcode `\_12\catcode `\%12\relax}%
\providecommand \@@startlink[1]{}%
\providecommand \@@endlink[0]{}%
\providecommand \url  [0]{\begingroup\@sanitize@url \@url }%
\providecommand \@url [1]{\endgroup\@href {#1}{\urlprefix }}%
\providecommand \urlprefix  [0]{URL }%
\providecommand \Eprint [0]{\href }%
\@ifxundefined \urlstyle {%
  \providecommand \doi  [0]{\begingroup \@sanitize@url \@doi}%
  \providecommand \@doi [1]{\endgroup \@@startlink {\doibase
  #1}doi:\discretionary {}{}{}#1\@@endlink }%
}{%
  \providecommand \doi  [0]{doi:\discretionary{}{}{}\begingroup
  \urlstyle{rm}\Url }%
}%
\providecommand \doibase [0]{http://dx.doi.org/}%
\providecommand \Doi [0]{\begingroup \@sanitize@url \@Doi }%
\providecommand \@Doi  [1]{\endgroup\@@startlink{\doibase#1}\@@Doi}%
\providecommand \@@Doi [1]{#1\@@endlink}%
\providecommand \selectlanguage [0]{\@gobble}%
\providecommand \bibinfo  [0]{\@secondoftwo}%
\providecommand \bibfield  [0]{\@secondoftwo}%
\providecommand \translation [1]{[#1]}%
\providecommand \BibitemOpen [0]{}%
\providecommand \bibitemStop [0]{}%
\providecommand \bibitemNoStop [0]{.\EOS\space}%
\providecommand \EOS [0]{\spacefactor3000\relax}%
\providecommand \BibitemShut  [1]{\csname bibitem#1\endcsname}%
\end{thebibliography}%


\begin{thebibliography}{}

\bibitem[Gerlich, 1992]{gerlich92}
Gerlich, D. 1992.
\newblock Inhomogeneous rf fields: a versatile tool for the study of processes
  with slow ions.
\newblock Advances in Chemical Physics Series, 82.

\bibitem[Wester, 2009]{wester09}
Wester, Roland 2009.
\newblock Radiofrequency multipole traps: tools for spectroscopy and dynamics
  of cold molecular ions.
\newblock Journal of Physics B: Atomic, Molecular and Optical Physics,
  42(15):154001 (12pp).
  
\bibitem[Prestage \& Weaver, 2007]{prestage07}
Prestage, J.D., \& G.L. Weaver 2007.
\newblock Atomic Clocks and oscillators for Deep-Space navigation and Radio
  Science.
\newblock Proceedings of the IEEE, 95(11):2235 -- 2247.

\bibitem[Okada et~al., 2007]{okada07}
Okada, Kunihiro, Kazuhiro Yasuda, Toshinobu Takayanagi, Michiharu Wada, Hans~A.
  Schuessler, \& Shunsuke Ohtani 2007.
\newblock Crystallization of Ca$^+$ ions in a linear rf octupole ion trap.
\newblock Phys. Rev. A, 75(3):033409.

\bibitem[Okada et~al., 2009]{okada09}
Okada, K., T.~Takayanagi, M.~Wada, S.~Ohtani, \& H.~A. Schuessler 2009.
\newblock Observation of ion Coulomb crystals in a cryogenic linear octupole rf
  ion trap.
\newblock Phys. Rev. A, 80(4):043405.

\bibitem[Champenois, 2009]{champenois09}
Champenois, C 2009.
\newblock About the dynamics and thermodynamics of trapped ions.
\newblock Journal of Physics B: Atomic, Molecular and Optical Physics,
  42(15):154002 (9pp).

\bibitem[Calvo et~al., 2009]{calvo09}
Calvo, F., C.~Champenois, \& E.~Yurtsever 2009.
\newblock Crystallization of ion clouds in octupole traps: Structural
  transitions, core melting, and scaling laws.
\newblock Phys. Rev. A, 80(6):063401.

\bibitem[Yurtsever et~al., 2011]{yurtsever11}
Yurtsever, E., E.~D. Onal, \& F.~Calvo 2011.
\newblock Structure and dynamics of ion clusters in linear octupole traps:
  Phase diagrams, chirality, and melting mechanisms.
\newblock Phys. Rev. A, 83(5):053427.

\bibitem[Champenois et~al., 2010]{champenois10}
Champenois, C., M.~Marciante, J.~Pedregosa-Gutierrez, M.~Houssin, M.~Knoop, \&
  M.~Kajita 2010.
\newblock Ion ring in a linear multipole trap for optical frequency metrology.
\newblock Phys. Rev. A, 81(4):043410.

\bibitem[Marciante et~al., 2010]{marciante10}
Marciante, M., C.~Champenois, A.~Calisti, J.~Pedregosa-Gutierrez, \& M.~Knoop
  2010.
\newblock Ion dynamics in a linear radio-frequency trap with a single cooling
  laser.
\newblock Phys. Rev. A, 82(3):033406.

\bibitem[Marciante et~al., 2011]{marciante11}
Marciante, M., C.~Champenois, J.~Pedregosa-Gutierrez, A.~Calisti, \& M.~Knoop
  2011.
\newblock Parallel ion strings in linear multipole traps.
\newblock Phys. Rev. A, 83(2):021404.

\bibitem[Prestage et~al., 1991]{prestage91b}
Prestage, J.~D., A.~Williams, L.~Maleki, M.~J. Djomehri, \& E.~Harabetian 1991.
\newblock Dynamics of charged particles in a Paul radio-frequency quadrupole
  trap.
\newblock Phys. Rev. Lett., 66(23):2964--2967.

\bibitem[Schiffer et~al., 2000]{schiffer00}
Schiffer, J.P., M.~Drewsen, J.S. Hangst, \& L.~Hornek{\ae}r 2000.
\newblock Temperature, ordering,and equilibrium with time-dependent confining
  force.
\newblock PNAS, 97:10697.

\bibitem[Otto et~al., 2009]{otto09}
Otto, R., P.~Hlavenka, S.~Trippel, J.~Mikosch, K.~Singer, M.~Weidemueller, \&
  R.~Wester 2009.
\newblock How can a 22-pole ion trap exhibit 10 local minima in the effective
  potential?
\newblock J. Phys. B, 42:154007.

\bibitem[Waki et~al., 1992]{waki92}
Waki, I., S.~Kassner, G.~Birkl, \& H.~Walther 1992.
\newblock Observation of ordered structures of laser-cooled ions in a
  quadrupole storage ring.
\newblock Phys. Rev. Lett., 68(13):2007--2010.

\bibitem[Ryjkov et~al., 2005]{ryjkov05}
Ryjkov, Vladimir~L., XianZhen Zhao, \& Hans~A. Schuessler 2005.
\newblock Simulations of the rf heating rates in a linear quadrupole ion trap.
\newblock Phys. Rev. A, 71(3):033414.

\bibitem[Olmos et~al., 2009]{olmos09}
Olmos, B., R.~Gonz\'alez-F\'erez, \& I.~Lesanovsky 2009.
\newblock Fermionic Collective Excitations in a Lattice Gas of Rydberg Atoms.
\newblock Phys. Rev. Lett., 103(18):185302.

\bibitem[Wineland et~al., 1978]{wineland78}
Wineland, D.~J., R.~E. Drullinger, \& F.~L. Walls 1978.
\newblock Radiation-Pressure Cooling of Bound-Resonant Absorbers.
\newblock Phys. Rev. Lett., 40(25):1639.

\bibitem[Mikosch et~al., 2007]{mikosch07}
Mikosch, J., U.~Fr\"{u}hling, S.~Trippel, D.~Schwalm, M.~Weidem\"{u}ller, \&
  R.~Wester 2007.
\newblock Evaporation of Buffer-Gas-Thermalized Anions out of a Multipole rf
  Ion Trap.
\newblock Phys. Rev. Lett., 98(22):223001.

\bibitem[Mikosch et~al., 2008]{mikosch08}
Mikosch, J., U.~Fr\"{u}hling, S.~Trippel, R.~Otto, P.~Hlavenka, D.~Schwalm,
  M.~Weidem\"{u}ller, \& R.~Wester 2008.
\newblock Evaporation of trapped anions studied with a 22-pole ion trap in
  tandem time-of-flight configuration.
\newblock Phys. Rev. A, 78(2):023402.

\end{thebibliography}

\end{document}